\documentclass[11pt]{article}
\usepackage{latexsym}
\usepackage{graphicx}
\usepackage{amsfonts,amssymb}

\begin{document}

\title{Quantum Prisoner's Dilemma in the new representation}
\author{Jinshan Wu \\
Department of Physics, Simon Fraser University, Burnaby, B.C. Canada, V5A 1S6%
}
\maketitle

\begin{abstract}
Using the representation introduced in our another
paper\cite{frame}, the well-known Quantum Prisoner's Dilemma
proposed in \cite{jens}, is reexpressed and calculated. By this
example and the works in \cite{frame} on classical games and
Quantum Penny Flip game, which first proposed in \cite{meyer}, we
show that our new representation can be a general framework for
games originally in different forms.
\end{abstract}

Key Words: Quantum Game Theory, Prisoner's Dilemma

Pacs: 02.50.Le, 03.67.-a

\textit{Introduction} --- Recently, we proposed a new mathematical
representation\cite{frame} for Classical and Quantum Game Theory.
It has been shown than $N$-player classical games, which are
traditionally defined by $N$ single-player strategy sets and
$\left(0,N\right)$-tensor payoff functions, can be equivalently
reexpressed in the new representation, by a system strategy space
and $N$ $(1,1)$-tensor payoff functions. In the same paper, the
well-known Quantum Penny Flip game\cite{meyer} has also been
rewritten by the new language. In another paper\cite{entangle}, we
apply the new representation onto Battle of the Sexes and get some
interesting results such as entangled strategy equilibrium state.
Although our new representation is defined as an abstract form,
which is believed be able to describe any specific games, it seems
still necessary to discuss more famous games as examples by this
new language. So in this paper, we try to describe in the new
representation a well-known game proposed in \cite{jens}, the
Quantum Prisoner's Dilemma.

{\it{The original Quantum Prisoner's Dilemma}} --- First, we
follow the definition in \cite{jens}, but reexpress it in density
matrix form instead of the original state vector form, and give
the manipulative definition proposed in \cite{nash},
\begin{equation}
\Gamma^{q,o} = \left(\rho^{q}_{0}\in\mathbb{H}^{q},
\prod_{i=1}^{N}\otimes \mathbb{H}^{i}, \mathcal{L},
\left\{P^{i}\right\}\right). \label{qogame}
\end{equation}
A two-particle quantum system is used as the quantum object in the
game, which has the Hilbert space formed by base vectors
$\left|UU\right>,\left|UD\right>,\left|DU\right>,\left|DD\right>$.
Here $\left|UU\right>$ represents the state that particle $1$ and
particle $2$ stay on $\left|U\right>$. We also suppose they are
distinguishable, named $1$ and $2$ respectively. The initial state
of the quantum object is
\begin{equation}
\rho^{q}_{0} = \left|UU\right>\left<UU\right| =
\left[\begin{array}{cccc}1 & 0 & 0 & 0\\0 & 0 & 0 & 0\\0 & 0 & 0 &
0\\0 & 0 & 0 & 0\end{array}\right].
\end{equation}
A typical strategy players can use is
\begin{equation}
\hat{U}(\theta, \phi) =
\left[\begin{array}{cc}e^{i\phi}\cos\theta/2 & \sin\theta/2
\\-\sin\theta/2 & e^{-i\phi}\cos\theta/2
\end{array}\right].
\label{quantumstrategy}
\end{equation}
So player $1$'s strategy space is the above unitary operator
acting on particle $1$, and similarly for player $2$. The payoff
value is determined by
\begin{equation}
E^{i} =
Tr\left(G^{i}\left(\hat{U}^{1}\otimes\hat{U}^{2}\right)\rho_{0}\left(\hat{U}^{1}\otimes\hat{U}^{2}\right)^{\dag}\right),
\label{jenspayoff}
\end{equation}
in which, if defined in the base vectors above, the payoff scale
matrix $P^{i}$ is
\begin{equation}
\begin{array}{ccc}
P^{1} = \left[\begin{array}{cccc}r & 0 & 0 & 0
\\0 & s & 0 & 0
\\0 & 0 & t & 0
\\0 & 0 & 0 & p\end{array}\right] & \mbox{and} & P^{2} = \left[\begin{array}{cccc}r & 0 & 0 & 0
\\0 & t & 0 & 0
\\0 & 0 & s & 0
\\0 & 0 & 0 & p\end{array}\right].
\end{array}
\end{equation}
So mapping $\mathcal{L}\left(\hat{U}^{1},\hat{U}^{2}\right) =
\hat{U}^{1}\otimes\hat{U}^{2}$. Classical pure strategies are
\begin{equation}
\begin{array}{ccc}
N^{c} = \left[\begin{array}{cc}1 & 0
\\0 & 1 \end{array}\right] & \mbox{and} & F^{c} = \left[\begin{array}{cc}0 & 1
\\1 & 0\end{array}\right].
\end{array}
\label{classicalbase}
\end{equation}
If players can only use classical strategies, we can check that
the payoff from equ(\ref{jenspayoff}) equal to the classical
payoff defined as
\begin{equation}
\begin{array}{ccc}
G^{1,c} = \left[\begin{array}{cc}r & s
\\t & p \end{array}\right] & \mbox{and} & G^{2,c} = \left[\begin{array}{cc}r &
t
\\s & p\end{array}\right].
\end{array}
\label{classicalpayoff}
\end{equation}
For example, we check the situation when both the two players
choose $N^{c}$. $N^{c}$ acting on $\left|U\right>$ gives
$\left|U\right>$, so the end state of the quantum object is still
$\left|UU\right>$. So the first elements of $G^{1,c}$ and
$G^{2,c}$ are $r$ and $r$.

{\it{Quantum Prisoner's Dilemma in the new representation}} ---
Now we try to derive the abstract form\cite{frame,nash},
\begin{equation}
\Gamma^{q} = \left(\prod_{i=1}^{N}\otimes S^{q}_{i},
\left\{H^{i}\right\}\right). \label{qgame}
\end{equation}
The central idea of our new representation is to find a set of
base vectors for strategy, and to defined inner product between
them so as to form them as a Hilbert space. And then redefined
payoff function as a mapping from the system strategy space to
real number. Here, we have four good and natural base strategies.
Besides the two classical pure strategies in
equ(\ref{classicalbase}), we still have
\begin{equation}
\begin{array}{ccc}
N^{q} = \left[\begin{array}{cc}1 & 0
\\0 & -1 \end{array}\right] & \mbox{and} & F^{q} = \left[\begin{array}{cc}0 &
-i
\\i & 0\end{array}\right],
\end{array}
\label{quantumbase}
\end{equation}
which we named as quantum base strategies. A general quantum
strategy in equ(\ref{quantumstrategy}) can be expanded as
\begin{equation}
\hat{U}\left(\theta, \phi\right) = \cos\frac{\theta}{2}\cos\phi
N^{c} + i\cos\frac{\theta}{2}\sin\phi N^{q} +
i\sin\frac{\theta}{2} F^{q}.
\end{equation}
A more general operator can be
\begin{equation}
\hat{U} = \xi\cdot N^{c} + x \cdot F^{c} + y \cdot F^{q} + z \cdot
N^{q}, \left(\xi,x,y,z \in \mathbb{C}\right).
\label{generalstatevector}
\end{equation}
Or if we require $s$ is unitary, in a set of independent
parameters\cite{qit},
\begin{equation}
\hat{U} = \cos\frac{\gamma}{2}\cos\frac{\alpha+\beta}{2}N^{c}
-i\sin\frac{\gamma}{2}\sin\frac{\alpha-\beta}{2}F^{c}
-i\sin\frac{\gamma}{2}\cos\frac{\alpha-\beta}{2}F^{q}
-i\cos\frac{\gamma}{2}\sin\frac{\alpha+\beta}{2}N^{q}.
\label{generalunitarystate}
\end{equation}
The inner product is defined as
\begin{equation}
\left(s,s^{'}\right) =
\frac{Tr\left(s^{\dag}s^{'}\right)}{Tr\left(I\right)}.
\end{equation}
Then $\left(N^{c}, F^{c}, N^{q}, F^{q}\right)$ are orthogonal and
normalized. Later on we denote them as base vectors such as
$\left|N^{c}\right>$. A system strategy space is the direct
product space of the two players, so it has $16$ base vectors such
as $\left|N^{c},N^{c}\right>$. A state in the system strategy
space can be
\begin{equation}
\left|S\right> = \left|s^{1}, s^{2}\right>.
\end{equation}
Now we try to define $\left(1,1\right)$-tensor payoff matrix
$H^{i}$ so that
\begin{equation}
E^{i}\left(S\right) = \left<S\right|H^{i}\left|S\right>, \forall
S.
\label{newpayoffvalue}
\end{equation}
In \cite{frame}, a general procedure has been proposed, that first
to define its elements on a specific set of base vectors, then
prove it can be used for any states. Now elements of the payoff
matrix is defined
\begin{equation}
H^{i}_{\alpha\beta} = \left<\alpha\right|H^{i}\left|\beta\right> =
\left<\alpha^{1},
\alpha^{2}\right|H^{i}\left|\beta^{1},\beta^{2}\right> =
Tr\left(P^{i}\left(\beta^{1}\otimes\beta^{2}\right)\rho_{0}\left(\alpha^{1}\otimes\alpha^{2}\right)^{\dag}\right),
\label{newpayoff}
\end{equation}
in which $\alpha^{i},\beta^{i}$ are anyone of the predefined base
vectors $\left(N^{c}, F^{c}, N^{q}, F^{q}\right)$. Before we
calculate all the values of the elements, we need to prove the
definition in equ(\ref{newpayoff}) guarantee
equ(\ref{newpayoffvalue}) is valid for any strategy.

\noindent {\bf{Theorem}} Suppose $\left|S\right>
=\left|s^{1},s^{2}\right>$, $\forall s^{1},s^{2}$, for the payoff
matrix $H^{i}$ defined in equ(\ref{newpayoff}), prove that
$E^{i}\left(S\right) = Tr\left(P^{i}\left(s^{1}\otimes
s^{2}\right)\rho_{0}\left(s^{1}\otimes s^{2}\right)^{\dag}\right)$
equals $\left<S\right|H^{i}\left|S\right> = \left<s^{1},
s^{2}\right|H^{i}\left|s^{1},s^{2}\right>$.

\noindent {\bf{Proof}} If $s^{1},s^{2}$ are the base vectors, this
is just the definition of $H^{i}$. So it's obvious. We claim that
$\left(s^{1}\otimes s^{2}\right)^{\dag} =
\left(s^{1}\right)^{\dag}\otimes \left(s^{2}\right)^{\dag}$,
$x_{\alpha} \cdot s^{1}_{\alpha}\otimes y_{\nu} \cdot s^{2}_{\nu}
= x_{\alpha} y_{\nu} \cdot s^{1}_{\alpha}\otimes s^{2}_{\nu}$ and
$\left(s^{1}_{\alpha} + s^{1}_{\beta}\right)\otimes
\left(s^{2}_{\mu} + s^{2}_{\nu}\right)= s^{1}_{\alpha}\otimes
s^{2}_{\mu} + s^{2}_{\beta}\otimes s^{2}_{\mu} +
s^{1}_{\alpha}\otimes s^{2}_{\nu} + s^{1}_{\beta}\otimes
s^{2}_{\nu}$. The proof of a general strategy $S$ will need all of
these relations, which are easy to check. Now we suppose $s^{i} =
\sum_{\mu}x^{i}_{\mu}\left|\mu\right>$. Then
\[
\begin{array}{lll}
Tr\left(P^{i}\left(s^{1}\otimes
s^{2}\right)\rho_{0}\left(s^{1}\otimes s^{2}\right)^{\dag}\right)
& = & Tr\left(P^{i}\left(s^{1}\otimes
s^{2}\right)\rho_{0}\left(s^{1}\right)^{\dag}\otimes
\left(s^{2}\right)^{\dag}\right)
\\ & = & Tr\left(P^{i}\sum_{\mu,\nu}x^{1}_{\mu}x^{2}_{\nu}\mu\otimes
\nu\rho_{0}\left(\sum_{\xi}x^{1}_{\xi}\xi\right)^{\dag}\otimes
\left(\sum_{\eta}x^{2}_{\eta}\eta\right)^{\dag}\right)
\\& = & Tr\left(P^{i}\sum_{\mu,\nu}x^{1}_{\mu}x^{2}_{\nu}\mu\otimes
\nu\rho_{0}\sum_{\xi,\eta}\bar{x}^{1}_{\xi}\bar{x}^{2}_{\eta}\xi^{\dag}\otimes
\eta^{\dag}\right)
\\ & = & Tr\left(P^{i}\sum_{\mu,\nu,\xi,\eta}x^{1}_{\mu}x^{2}_{\nu}\bar{x}^{1}_{\xi}\bar{x}^{2}_{\eta}\mu\otimes
\nu\rho_{0}\xi^{\dag}\otimes \eta^{\dag}\right)
\\ & = & \sum_{\mu,\nu,\xi,\eta}x^{1}_{\mu}x^{2}_{\nu}\bar{x}^{1}_{\xi}\bar{x}^{2}_{\eta}Tr\left(P^{i}\mu\otimes
\nu\rho_{0}\xi^{\dag}\otimes \eta^{\dag}\right)
\\ & = &
\sum_{\mu,\nu,\xi,\eta}x^{1}_{\mu}x^{2}_{\nu}\bar{x}^{1}_{\xi}\bar{x}^{2}_{\eta}\left<\xi,\eta\left|H\right|\mu,\nu\right>
\end{array}
\]
and
\[
\begin{array}{lll}
\left<S\right|H^{i}\left|S\right> & = & \left<s^{1},
s^{2}\right|H^{i}\left|s^{1},s^{2}\right>
\\ & = & \sum_{\xi,\eta}\bar{x}^{1}_{\xi}\bar{x}^{2}_{\eta}\left<\xi\left|\left<\eta\left|H\sum_{\mu,\nu}x^{1}_{\mu}x^{2}_{\nu}\right|\mu\right>\right|\nu\right>
\\ & = & \sum_{\mu,\nu,\xi,\eta}x^{1}_{\mu}x^{2}_{\nu}\bar{x}^{1}_{\xi}\bar{x}^{2}_{\eta}\left<\xi,\eta\left|H\right|\mu,\nu\right>
\end{array}
\]
So they are equal, and we get equ(\ref{newpayoffvalue}). The
payoff matrix $H^{1}, H^{2}$ in the base vectors $\left(N^{c},
F^{c}, N^{q}, F^{q}\right)$ are
\begin{equation}
H^{1} = \left[
\begin{array}{cccccccccccccccc}
r & 0 & r & 0 & 0 & 0 & 0 & 0 & r & 0 & r & 0 & 0 & 0 & 0 & 0 \\
0 & s & 0 & is & 0 & 0 & 0 & 0 & 0 & s & 0 & is & 0 & 0 & 0 & 0 \\
r & 0 & r & 0 & 0 & 0 & 0 & 0 & r & 0 & r & 0 & 0 & 0 & 0 & 0 \\
0 & -is & 0 & s & 0 & 0 & 0 & 0 & 0 & -is & 0 & s & 0 & 0 & 0 & 0 \\
0 & 0 & 0 & 0 & t & 0 & t & 0 & 0 & 0 & 0 & 0 & it & 0 & it & 0 \\
0 & 0 & 0 & 0 & 0 & p & 0 & ip & 0 & 0 & 0 & 0 & 0 & ip & 0 & -p \\
0 & 0 & 0 & 0 & t & 0 & t & 0 & 0 & 0 & 0 & 0 & it & 0 & it & 0 \\
0 & 0 & 0 & 0 & 0 & -ip & 0 & p & 0 & 0 & 0 & 0 & 0 & p & 0 & ip \\
r & 0 & r & 0 & 0 & 0 & 0 & 0 & r & 0 & r & 0 & 0 & 0 & 0 & 0 \\
0 & s & 0 & is & 0 & 0 & 0 & 0 & 0 & s & 0 & is & 0 & 0 & 0 & 0 \\
r & 0 & r & 0 & 0 & 0 & 0 & 0 & r & 0 & r & 0 & 0 & 0 & 0 & 0 \\
0 & -is & 0 & s & 0 & 0 & 0 & 0 & 0 & -is & 0 & s & 0 & 0 & 0 & 0 \\
0 & 0 & 0 & 0 & -it & 0 & -it & 0 & 0 & 0 & 0 & 0 & t & 0 & t & 0 \\
0 & 0 & 0 & 0 & 0 & -ip & 0 & p & 0 & 0 & 0 & 0 & 0 & p & 0 & ip \\
0 & 0 & 0 & 0 & -it & 0 & -it & 0 & 0 & 0 & 0 & 0 & t & 0 & t & 0 \\
0 & 0 & 0 & 0 & 0 & -p & 0 & -ip & 0 & 0 & 0 & 0 & 0 & -ip & 0 & p%
\end{array}\right],
\end{equation}
and
\begin{equation}
H^{2} = H^{1}\left(t\rightleftarrows s\right).
\end{equation}
For classical game, the base vectors are only $N^{c}, F^{c}$, then
the sub-matrix are
\begin{equation}
\begin{array}{ccc}
H^{1,c} = \left[\begin{array}{cccc}r & 0 & 0 & 0 \\0 & s & 0 & 0
\\0 & 0 & t & 0 \\0 & 0 & 0 & p
\end{array}\right] & \mbox{and} & H^{2,c} = \left[\begin{array}{cccc}r & 0 & 0 & 0 \\0 & t & 0 & 0 \\0 & 0 & s & 0 \\0 & 0 & 0 & p
\end{array}\right].
\end{array}
\end{equation}
They are equivalent with the payoff matrix directly reexpressed
into our new representation from $G^{1,c}, G^{2,c}$.

{\it{Density matrix form of the game}} --- For a quantum system
state, equ(\ref{newpayoffvalue}) can be used to calculated the
payoff value. But a classical mixture strategy with probability
$p^{i}_{nc}$ on $N^{c}$ and $p^{i}_{fc}$ on $F^{c}$ is impossible
to rewritten as the vector form as equ(\ref{generalstatevector}).
In order to compare quantum strategy with classical strategy, we
have to define a more general strategy form. In \cite{frame}, a
density matrix form is used, such as
\[
\rho^{S,c} =\left(p^{1}_{nc}\left|N^{c}\right>\left<N^{c}\right| +
p^{1}_{nc}\left|N^{c}\right>\left<N^{c}\right|\right)\left(p^{2}_{nc}\left|N^{c}\right>\left<N^{c}\right|
+ p^{2}_{nc}\left|N^{c}\right>\left<N^{c}\right|\right).
\]
In fact, this density matrix form can be applied onto both
classical and quantum strategies. So quantum mixture strategy is
permitted to use by a quantum player. Then the payoff value
equ(\ref{newpayoffvalue}) turns into a density matrix form,
\begin{equation}
E^{i} = Tr\left(\rho^{S}H^{i}\right).
\end{equation}
Now our classical and quantum Prisoner's Dilemma is redefined as
\begin{equation}
\begin{array}{ccc}
\Gamma^{q} = \left(S^{1,q}\otimes
S^{2,q},\left(H^{1,q},H^{2,q}\right) \right) & \mbox{and} &
\Gamma^{c} = \left(S^{1,c}\otimes
S^{2,c},\left(H^{1,c},H^{2,c}\right) \right)
\end{array}.
\end{equation}
The classical game is defined in a subspace of the quantum game,
and the classical payoff matrix is the sub-matrix on the subspace.

{\it{Equilibrium state of the game}} --- Now we have shown that
our language can be used to discuss this game. Although
calculation of NE and a general algorithm is not the main topic of
this paper, finding some solutions and comparing them with the
solutions given in their original frameworks is quite attractive.
In \cite{entangle}, Nash Equilibrium is redefined and a Global
Equilibrium State (GES) is proposed. And in \cite{jens}, a Pareto
optimal state $\left(\hat{U}\left(0,\pi/2\right),
\hat{U}\left(0,\pi/2\right)\right)$ is found. Now we try to check
if there is a GES, if not, if there is some other state which can
be used to beat the Parato optimal strategy. A Nash Equilibrium
State $\rho^{S}_{eq}$ is defined as
\begin{equation}
E^{i}\left(\rho_{eq}^{s}\right) \geq
E^{i}\left(Tr^{i}\left(\rho_{eq}^{s}\right)\cdot\rho^{i}\right),
\forall i, \forall \rho^{i}, \label{extendedequilibrium}
\end{equation}
in which $Tr^{i}\left(\cdot\right)$ means to do the trace in
player $i$'s strategy space. If system state is a direct product
of all single-player states,
\begin{equation}
\left(\rho_{eq}^{s}\right) = \prod_{i}\rho_{eq}^{i}
\end{equation}
then definition in equ(\ref{extendedequilibrium}) is equivalent
with traditional NE,
\begin{equation}
E^{i}\left(\rho_{eq}^{s}\right) \geq
E^{i}\left(\rho_{eq}^{1}\cdots\rho^{i}\cdots\rho_{eq}^{N}\right),
\forall i, \forall \rho^{i}. \label{NE}
\end{equation}
A special case of the first definition is
\begin{equation}
E^{i}\left(\rho_{eq,m}^{s}\right) \geq E^{i}\left(\rho^{s}\right),
\forall \rho^{s}, \forall i.  \label{specialcase}
\end{equation}
Although it is not always possible to find such a state
$\rho^{S}_{eq,m}$, if the game has one such state then it is a
dominant strategy, so we named it GES\cite{entangle}. The reduced
payoff matrix $H^{i}_{R}$ is the reduced matrix of $H^{i}$ when
all other players' strategies are fixed,
\begin{equation}
H_{R}^{i} =
Tr_{-i}(\rho^{1}\cdots\rho^{i-1}\rho^{i+1}\cdots\rho^{N}H^{i}),
\label{reducedpayoff}
\end{equation}
where $Tr_{-i}\left(\cdot\right)$ means to do the trace in the
space except player $i$'s space.

If both $H^{1}$ and $H^{2}$ have a common eigenvector, which has
the maximum eigenvalue in both the two payoff matrix, then this
state is a GES. It's probably an entangled strategy state like the
one of the game in \cite{entangle}. Here we check if such state
exists in this game. Both $H^{1}$ and $H^{2}$ have eigenvalues
$4s, 4r, 4p, 4t$ and other $12$ zeros. The corresponding
eigenvector $\left|4t\right>^{1}\neq\left|4t\right>^{2}$ and
$\left|4s\right>^{1}\neq\left|4s\right>^{2}$, but
$\left|4r\right>^{1} = \left|4r\right>^{2}$ and
$\left|4p\right>^{1} = \left|4p\right>^{2}$. When $t>r>p>s$, there
is no GES, but $\left|S_{m}\right> = \left|4r\right>^{1} =
\left|4r\right>^{2}$ are the system state with second-maximum
eigenvalue, on which both player $1$ and player $2$ get $4r$. The
vector form of $\left|S_{m}\right>$ is
\[
\left|S_{m}\right> = \frac{1}{2}\left(\left|N^{c}N^{c}\right> +
\left|N^{c}N^{q}\right> + \left|N^{q}N^{c}\right> +
\left|N^{q}N^{q}\right> \right) =
\frac{1}{2}\left(\left|N^{c}\right> +
\left|N^{q}\right>\right)\left(\left|N^{c}\right> +
\left|N^{q}\right> \right),
\]
or transfer it back into matrix form
\[
S_{m} = \left[\begin{array}{cccc}2 & 0 & 0 & 0 \\ 0 & 0 & 0 & 0 \\
0 & 0 & 0 & 0 \\ 0 & 0 & 0 & 0
\end{array}\right] = \left[\begin{array}{cc}\sqrt{2} & 0 \\ 0 & 0
\end{array}\right]\otimes\left[\begin{array}{cc}\sqrt{2} & 0 \\ 0 & 0
\end{array}\right].
\]
So $S_{m}$ is not a unitary operator, although it leads to higher
payoff, it might be unapplicable. And even it's applicable, it's
not a NE. Because, player $i$ can get more payoff by adjust its
own strategy. The role of such system state is that everyone knows
it's not a best choice, but a good choice if both players can keep
staying on such state, just like the $\left|N^{c}N^{c}\right>$
state in classical prisoner's dilemma. Leaving from such state
will at least decrease the payoff of one player. Also state
$\left|4p\right>^{1} = \left|4p\right>^{1}$ has such similar
property.

Now we discuss the reduced payoff matrix when player $2$ or $1$ choose $\hat{%
U}\left( \theta_{2},\phi_{2}\right) $. From
equ(\ref{reducedpayoff}),
\begin{equation}
H_{R}^{1} = \left[
\begin{array}{cccc}
\epsilon_{1} & 0 & \epsilon_{1} & 0 \\
0 & \epsilon_{2} & 0 & i\epsilon_{2}  \\
\epsilon_{1}  & 0 & \epsilon_{1}  & 0 \\
0 & -i\epsilon_{2} & 0 & \epsilon_{2}
\end{array}
\right].
\end{equation}%
in which $\epsilon_{1} = s\sin ^{2}\frac{1}{2}\theta_{2} +r\cos
^{2}\frac{1}{2}\theta_{2}, \epsilon_{2} = p\sin
^{2}\frac{1}{2}\theta_{2} +t\cos ^{2}\frac{1}{2}\theta_{2}$. It
has eigenvalues $\left\{2\epsilon_{2},2\epsilon_{1},0,0\right\} $.
When $t>r>p>s$, the $2\epsilon_{2}$ is the maximum eigenvalue for
any $\theta _{2}$. The corresponding eigenvector is $\left(
0,i,0,1\right) ^{T}$, or in matrix form,
\[
s_{m}^{1}=\left[
\begin{array}{cc}
0 & 0 \\
\sqrt{2}i & 0%
\end{array}%
\right] ,
\]%
which is obviously not a unitary matrix. Although it leads to
higher payoff, it's not applicable. The same situation happens to
player $2$, so the payoff value of the both players will be $4p$.
So non-unitary operators space gives a new NE,
$\left|E^{1}=4p=E^{2}\right>$.

Now we limit out strategies in unitary operator space. For a
general unitary operator strategy $\hat{U}\left(
\theta_{1},\phi_{1}\right)$, the payoff of player $i$ is
\begin{equation}
E^{i} = \left(t\cos ^{2}\frac{\theta_{\left(3-i\right)} }{2}+p\sin
^{2}\frac{\theta_{\left(3-i\right)}
}{2}\right)\sin^{2}\frac{\theta_{i}}{2} + \left(r\cos
^{2}\frac{\theta_{\left(3-i\right)}}{2} + s\sin
^{2}\frac{\theta_{\left(3-i\right)}}{2}\right)\cos
^{2}\frac{\theta_{i}}{2}. \label{specpayoffvalue}
\end{equation}
Since the first term is larger, the best response is $\theta_{1} =
\pi$. Similarly, when player $1$ is fixed at $\hat{U}\left(
\theta_{1},\phi_{1}\right)$. the best response of player $2$ is
$\hat{U}\left(\pi,\phi_{2}\right) = iF^{q}$. Therefor, the NE in
quantum unitary strategy is
\begin{equation}
\left|S\right> = \left|E^{1}=p,E^{2}=p\right> =
\left(i\left|F^{q}\right>i\left|F^{q}\right>\right).
\end{equation}
However, this NE strongly depends on equ(\ref{quantumstrategy}),
because we have more unitary operators. In the whole unitary
operator space defined by equ(\ref{generalunitarystate}), with the
parameters $\alpha,\beta,\gamma$, the payoff of player $i$ is
still in the form of equ(\ref{specpayoffvalue}), in which
$\theta_{i}$ is replaced with $\gamma_{i}$. So it's still
independent of $\alpha,\beta$. Therefor, NE in this whole unitary
operator space is
\begin{equation}
\left|E^{1}=p,E^{2}=p\right> =
\left|\hat{U}\left(\gamma=\pi,\alpha_{1},\beta_{1}\right),\hat{U}\left(\gamma=\pi,\alpha_{2},\beta_{2}\right)\right>,
\end{equation}
where in operator form,
\begin{equation}
\hat{U}\left(\gamma=\pi,\alpha_{1},\beta_{1}\right) =
-i\sin\frac{\alpha-\beta}{2}F^{c} +
i\sin\frac{\alpha+\beta}{2}F^{q}.
\end{equation}

So for our Quantum Prisoner's Dilemma, the payoff from Quantum
Nash Equilibrium is the same with the payoff from original
Classical Nash Equilibrium. In this sense, we say the quantization
does not solve the dilemma, although instead of only one NE in
classical case, here the quantized game has more NEs
($\forall\alpha,\beta$). But if non-unitary operators are
applicable, we will have better NEs, such as
$\left|E^{1}=4p=E^{2}\right>$ and $\left|E^{1}=4r=E^{2}\right>$.
And further more, if entangled states are permitted, the former
NEs will not be NEs anymore. In some cases, even GES can be
found\cite{entangle}. Usually, such GES will not be a direct
product state, so it includes correlation between players. This
property looks like cooperative behavior. Although this game has
no GES, it's still probably to find general NE as
equ(\ref{extendedequilibrium}) in this entangled strategy space.
Unfortunately, we have no applicable algorithm for such general
NE. But if we can find them, it will probably have bigger payoff
and also be unitary in the system space. Then, we can say, a
general NE in entangled strategy space solves the dilemma.
However, of course, since it't not a direct product state, it
implies that something like negotiation and agreement are the real
reason to solve the dilemma. Anyway, even that, it's a good news,
which means our representation is hopefully a way from
non-cooperative game to cooperative game.

{\it{Conclusion}} --- Now we see the new representation can be
applied onto the Quantum Prisoner's Dilemma. In classical strategy
space, it gives the same result with traditional language,
$\left|E^{1}=p=E^{2}\right>$; in quantum unitary operator strategy
space, some new NEs, which have the same payoff with classical NE,
appears, $\left|E^{1}=p=E^{2}\right>$; in general quantum strategy
space including non-unitary operators, two new NEs appear in such
game, $\left|E^{1}=4p=E^{2}\right>$ and
$\left|E^{1}=4r=E^{2}\right>$, but they are non-unitary operators;
and at last, if entangled strategy is permitted, the game here has
no GES, but still probably has NE. Unfortunately, now we have no
way to get such NE. The existence of NE in such strategy space
calls more investigation. In fact, the definition proposed here is
for general NE in any strategy space, but without a proof of the
existence and no applicable algorithm. And frankly, we even have
no idea if such general NE is meaningful or not, because it
requires non-direct-product state and/or non-unitary operator.
However, the point is no matter whether they have applicable
meaning or not, questions in Game Theory can be discussed in our
new representation. Hopefully, one day, it will bring new stuff
into Game Theory. And it should be able to prove that for all
linear-probability-combination classical game and all
linear-and-anti-linear-amplitude-combination quantum game, the new
representation is always valid\cite{nash}.

{\it{Acknowledgement}} --- The authors want to thank Dr. Shouyong
Pei and Zengru Di for their advices during the revision of this
paper. This work is partial supported by China NSF 70371072 and
70371073.

\end{document}